\pdfoutput=1
\documentclass[a4paper]{article}

\usepackage{itgspeech2023}    
\usepackage{times}            
\usepackage[english]{babel}   
\usepackage[ansinew]{inputenc}
\usepackage[T1]{fontenc}      
\usepackage[sort&compress,numbers]{natbib}	
\usepackage{amsmath,amssymb}
\usepackage{graphicx}
\usepackage[colorlinks=false,pdfborder={0 0 0}]{hyperref}
\usepackage{units}
\usepackage[acronym]{glossaries}

\makeglossaries
\usepackage{multirow}
\DeclareMathAlphabet{\mathcal}{OMS}{cmsy}{m}{n}

\usepackage{tikz}
\usepackage{pgfplots}
\usepgfplotslibrary{groupplots,dateplot}
\usepackage{nicefrac}

\usepackage{balance}

\newcommand{\vect}[1]{\ensuremath{\boldsymbol{\mathbf{#1}}}}

\DeclareMathOperator{\ExpOp}{\mathbb{E}}
\usetikzlibrary{positioning, shapes, calc, arrows}
\usepackage{stackengine}
\usepackage{booktabs}

\usepackage{tabularx}

\title{On Feature Importance and Interpretability of Speaker Representations}

\author{Frederik Rautenberg$^1$, Michael Kuhlmann$^1$, 
Jana Wiechmann$^2$, Fritz Seebauer$^2$, \\Petra Wagner$^2$, Reinhold Haeb-Umbach$^1$}

\address{$^1$\emph{\small Department of Communications Engineering, Paderborn University} \emph{\small Email: \{rautenberg, kuhlmann, haeb\}@nt.upb.de} \\
$^2$\emph{\small Phonetics Work Group, Bielefeld University}
\emph{\small Email: \{jana.wiechmann, fritz.seebauer, petra.wagner\}@uni-bielefeld.de }
  }

\newacronym{lime}{LIME}{Local Interpretable Model-Agnostic Explanations}
\newacronym{shap}{SHAP}{SHapley Additive exPlanations}
\newacronym{fvae}{FVAE}{Factorized Variational Autoencoder}
\newacronym{cca}{CCA}{Canonical Correlation Analysis}
\newacronym{nsc}{NSC}{Nautilus Speaker Characterization}
\newacronym{mse}{MSE}{Mean Squared Error}
\newacronym{svr}{SVR}{Support Vector Regressor}
\newacronym{svc}{SVC}{Support Vector Classifier}
\newacronym{lr}{LR}{Linear Regression}
\newacronym{vr}{VR}{Voting Regressor}
\newacronym{mlp}{MLP}{Multilayer Perceptron}
\newacronym{rbf}{RBF}{Radial Basis Function}
\newacronym{lda}{LDA}{Linear Discriminant Analysis}
\newacronym{rf}{RF}{Random Forest}
\newacronym{cpp}{CPPS}{Smoothed Cepstral Peak Prominence}
\newacronym{lmmse}{LMMSE}{Linear Minimum Mean Squared Error}
\newacronym{gap}{GAP}{Global Average Pooling}
\newacronym{cpc}{CPC}{Contrastive Predictive Coding}
\newacronym{vtlp}{VTLP}{Vocal Tract Length Perturbation}
\newacronym{lld}{LLD}{Low-Level Descriptor}
\newacronym{asr}{ASR}{automatic speech recognition}

\newacronym{eGeMAPS}{eGeMAPS}{Geneva Minimalistic Acoustic Parameter Set}
\sloppypar
\usepackage{array}
\usepackage{tabularx}
\newcolumntype{C}[1]{>{\centering\arraybackslash}p{#1}}

\newcolumntype{C}[1]{>{\centering\let\newline\\\arraybackslash\hspace{0pt}}m{#1}}

\begin{document}

\maketitle
\begin{abstract}
Unsupervised speech disentanglement aims at separating fast varying from slowly varying components of a speech signal. In this contribution, we take a closer look at the embedding vector representing the slowly varying signal components, commonly named the speaker embedding vector. We ask, which properties of a speaker's voice are captured and investigate to which extent do individual embedding vector components sign responsible for them, using the concept of Shapley values. Our findings show that certain speaker-specific acoustic-phonetic properties can be fairly well predicted from the speaker embedding, while the investigated more abstract voice quality features cannot. 
\end{abstract}

\section{Introduction}
A core concept of modern deep learning methods is to compute a vector representation of input data by means of a neural network. This concept is used for all kinds of data, be it images, text, audio or any other data. Even different modalities are  mapped to a common embedding vector space, which then allows for interactions among them, such that one modality can learn from the information present in the other. In this contribution, we look at embedding vectors that are meant to represent a speaker's voice properties. Those are used in various applications, such as speaker verification, \gls{asr}, diarization, or voice conversion. 

While most of those, such as the well-known d-vectors \cite{Varani2014_dvector} or x-vectors \cite{Snyder2018_xvector}, are learnt discriminatively using a speaker classification objective, there are also approaches that compute embedding vectors in a completely unsupervised fashion \cite{qian19_autovc, chou2019one, hsu2017unsupervised,  ebbers2021contrastive}. They disentangle a speech signal into two representations with the goal that one represents the speaker and the other the content. This can be achieved by a carefully designed bottleneck  \cite{qian19_autovc}, through appropriate normalization \cite{chou2019one}, or by mapping  short-term, i.e., fast variations of the signal, and temporally more stable signal components into different representations \cite{hsu2017unsupervised, ebbers2021contrastive}. The latter is done with the underlying hypothesis, that fast variations of the speech signal are caused by the linguistic content of what is spoken, while components that change at a much slower rate, are caused by speaker specific voice properties. Indeed, those approaches have been successfully used for voice conversion. Here the embedding vector, which represents the temporally stable parts of the signal, is replaced by that computed from a target speaker, before the speech signal is reconstructed \cite{sisman_2020_overview_vc}.

The purpose of this paper is to put this last hypothesis under verification. Calling the embedding vector that captures the temporally stable signal parts in the following ``speaker embedding vector'', we want to know what is actually represented in this vector. We are aware that the statistics of environmental noise are typically also temporally more stable than the speech signal, and thus will be cast to the speaker embedding vector, but this is not the focus of our work. We rather want to know, which voice properties are  stored in the representation, and even which components of the vector capture which property of a speaker's voice. We will consider both acoustic-phonetic features, such as average pitch, as well as voice quality categories, such as nasality or hoarseness. 

We used the following methodology: we train a machine learning model to predict the voice property under investigation from the speaker embedding vector. The accuracy of the predictor is a measure of how well this particular speaker property is represented in the embedding. We then determine which component of the input speaker embedding vector takes how much responsibility for the prediction result  and thus carries information about that property.
This is done by using the concept of \gls{shap} \cite{shapley1953value, lundberg2017unified}, a method from cooperative game theory. Feature importance as measured by Shapley values is a common tool in machine learning \cite{holzinger2022explainable}, but not yet so common in speech processing.  In \cite{sivasankaran21_interspeech} it was used to compute the attribution of input features to an estimated mask for speech enhancement, while in \cite{Markert2022_shapley_asr} Shapley values were employed to determine which features at the input are most influential for the \gls{asr} output.

While \gls{cca}, that was used in \cite{daga_paper} to calculate feature importance,  is restricted to linear models, Shapley values can be computed for any machine learning model. In our study, we employ more complex machine learning models to perform linear and non-linear classification or regression tasks on the speaker embedding vector. The investigations show that statistical averages of acoustic-phonetic features are represented fairly well in the speaker embedding vector, while the investigated voice quality factors nasality and hoarseness are not well represented, although these properties are considered to be temporally stable. This is an important finding, as it shows that the current speaker embedding vector extraction system is not well prepared for voice quality conversion. Using phonetic insights to identify the acoustic correlates of such voice quality categories can give directions how the estimation of the speaker vector should be modified to better capture these characteristics.

The paper is organized as follows: We first give a brief overview of the concept of Shapley values. Section~\ref{sec:setup} outlines our experimental setup, Section~\ref{sec:results} presents the results for predicting acoustic-phonetic and voice quality features, and Section~\ref{sec:conclusion} provides our conclusions.

\section{Shapley Feature Importance}
Assume we are given a fixed-dimensional speaker embedding vector. 
Our goal is to find out in which components of the embedding vector a certain acoustic or perceptual property of the signal is encoded.  To this end, we are going to train a machine learning model, a classifier or regressor, to predict that property from the speaker embedding vector at its input and then estimate the importance of each component of the input feature vector for that prediction. We compute the feature importance using Shapley values, a concept from cooperative game theory. Here, the machine learning model is the game and the components of the speaker embedding vector at its input are the players in the game. The prediction of the model is the payout of the game. In order to compute a fair distribution of the payout among the players, the Shapley value computes the average attribution of a player to the outcome. In our problem, a feature's Shapley value is a measure of its importance for the prediction for a specific input. By aggregating over all predictions, we obtain a feature vector component's average contribution to the model predictions.

Next, we briefly summarize the concept of Shapley values, and then describe an approximate computation, called \gls{shap}. We refer the reader to \cite{lundberg2017unified} for more details. We also show how SHAP simplifies in case of a linear model. 

\subsection{Shapley Values}

The key idea of computing a feature importance with Shapley values is to compare the model output with the feature in question being present at its input and with it being absent. However, because features can interact with each other, we need to consider every possible subset of features, called coalition. To be specific, let $\mathcal{X} = \left\{ x_1, \hdots, x_I  \right\}$ be a set of these features, whose elements also make up the feature vector $\vect{x} = (x_1, \ldots , x_I)^\mathrm{T}$, where both can be the input of the predictor.\footnote{We, here, at times use set notation, $\mathcal{X}$, and at times vector notation, $\vect{x}$, whatever is deemed more appropriate.} We now have a machine learning model $f(\cdot)$, whose output $\hat{y}=f(\mathcal{X})$ is the prediction for the label $y$ and let $x_i$ be one of the features, for which we want to compute the importance to that prediction. The contribution of a feature $x_i$ to the coalition $\mathcal{S}$ can be calculated with  
\begin{equation}
    \label{eq:diff_output}
  m_i(\mathcal{S}, x_i) = f(\mathcal{S} \cup \{x_i\}) - f(\mathcal{S}) \, ,
\end{equation}
where $\mathcal{S}\subseteq \mathcal{X} \backslash \{x_i\}$ is a subset without $x_i$ and the subset $\mathcal{S} \cup \{x_i\}$ that includes $x_i$. It is worthwhile to note that a model output with a feature absent can be approximated by replacing the feature by a random value, since, on average, random inputs should not be able to contribute to the prediction. This idea can be verified as follows: Noting that sampling is a Monte Carlo approximation to compute an expectation and assuming that the prediction of the machine learning model is given by $f(\mathcal{S} \cup \{x_i\})= \ExpOp_y[y|\mathcal{S} \cup \{x_i\}]$, which is the optimal prediction in the \gls{mse} sense. Averaging over $x_i$ gives
\begin{equation}
     \ExpOp_{x_i}\left[\ExpOp_y\left[y|\mathcal{S} \cup \{x_i\} \right]\right] = \ExpOp_y[y|\mathcal{S}] = f(\mathcal{S}) \, , 
\end{equation}
which is the model output with the feature $x_i$ being absent. To obtain the marginal contribution of the feature $x_i$, we compute a weighted average over all subsets $\mathcal{S}$, which results in the Shapley value of that feature
\begin{equation}
\label{Eq:shapley_values}
    \phi_i(f,\mathcal{X})= \sum_{\mathcal{S} \subseteq \mathcal{X} \backslash \{x_i\}} \frac{|\mathcal{S}|! \cdot \left(|\mathcal{X}| - |\mathcal{S}| -1\right)!}{|\mathcal{X}|!} \cdot m_i(\mathcal{S}, x_i) \, ,
\end{equation}
where $|\cdot|$ defines the cardinality of the set. The weighting is chosen such that subsets with only few and subsets with almost all features have a high weight, because they are particularly informative about the feature importance \cite{lundberg2017unified}. Eq.~\eqref{Eq:shapley_values} is a local explanatory model, i.e., it explains the  contribution of feature $x_i$ to the outcome for a particular feature vector $\vect{x}$. 
To obtain a global explanation, i.e., the Shapley values that explain the
contribution of the features to predictions from multiple observations, the Shapley values from each feature vector $\vect{x}_n$ for all observations $\vect{X}=(\vect{x}_1, \ldots , \vect{x}_N)$ have to be calculated and then averaged over all obtained values. Computing the Shapley values is computationally expensive, because the number of subsets grows exponentially with the number of features $|\mathcal{X}|$. However, methods that significantly reduce the complexity have been developed that approximate Eq.~\eqref{Eq:shapley_values}, one of them is briefly described next.

%
%
%
%

\subsection{SHapley Additive exPlanations}
\label{Sec:shap}
Building upon the insight that an explanation is a model by itself, however a simpler one than the original model to be explained, 
\gls{shap} \cite{lundberg2017unified} uses an additive  explanation model $g(\cdot)$ that locally approximates $f(\cdot)$: %
\begin{equation}
    \label{Eq:Additive_feature_attribution_methods}
    g(\mathbf{z}') = \phi_0 + \sum_{i=1}^I \phi_i \cdot z'_i \, ,
\end{equation}
where the original input feature $x_i$ is replaced by the simplified input $z_i\in\{0,1\}$, where $z_i=1$ corresponds to the feature being present and $z_i=0$ to the feature being absent. Further, $\phi_0$, is the approximate model output if no feature is at its input, which is given by $\phi_0 = \ExpOp_y[y]$.
In \cite{lundberg2017unified} it is shown that while several explanation models fall into this class of additive explanation models, only if the $\phi_i$ are the Shapley values, then certain desirable properties of explanation models are met. In the following we will thus use the explanation model of Eq.~\eqref{Eq:Additive_feature_attribution_methods}, with $\phi_i$ being the Shapley value for feature $x_i$.

\subsection{SHAP for Linear Models}
The linear minimum mean squared error estimator $\hat{y}$ of variable $y$, given an input feature vector $\vect{x}$, is given by 
\begin{equation}
    \label{eq:LMMSE}
    \hat{y}= f(\vect{x})= \ExpOp[y]+ \boldsymbol{\Sigma}_{\vect{x}y}^{\top}\boldsymbol{\Sigma}_{\vect{x}\vect{x}}^{-1}(\vect{x}-\ExpOp[\vect{x}])\,,
\end{equation}
where
$\boldsymbol{\Sigma}_{\vect{x}y} = \ExpOp[(\vect{x}-\ExpOp[\vect{x}])(y-\ExpOp[y])]$ is  the cross-covariance vector  with elements $\sigma_{x_iy}$, and $\boldsymbol{\Sigma}_{\vect{x}\vect{x}} =
\ExpOp[(\vect{x}-\ExpOp[\vect{x}])(\vect{x}-\ExpOp[\vect{x}])^\top]$ is the autocovariance matrix with elements $\sigma^2_{x_ix_j}$.
If the components of $\vect{x}$ are independent, then $\boldsymbol{\Sigma}_{\vect{x}\vect{x}}$ is a diagonal matrix with elements $\sigma^2_{x_ix_i}$, $i=1,\ldots I$ on the main diagonal. Thus, Eq.~\eqref{eq:diff_output}, the difference in the model's output with and without feature $x_i$,  is given by
\begin{equation}
    m_i(s_i) = \frac{\sigma_{x_iy}}{\sigma_{x_ix_i}^2} \, ,
\end{equation}
which is independent of the chosen coalition $\mathcal{S}$. Thus the weighting of different coalitions as in Eq.~\eqref{Eq:shapley_values} is unnecesary, and  $m_i(s_i)$
is equal to the Shapley value $\phi_i$.
We further note, that the linear model is identical to the \gls{cca} that has been used in \cite{daga_paper} to assess the feature importance for predicting a certain acoustic signal property $y$.

\section{Experimental Setup}
\label{sec:setup}

The goal of this investigation is to better understand which information is represented in a speaker embedding vector. Here, we analyze the speaker embedding that is computed with the \gls{fvae} that has been introduced in \cite{ebbers2021contrastive} and later used in \cite{Kuhlmann2022} for voice conversion.
In the following, we first briefly describe the \gls{fvae}, then describe the machine learning models for predicting acoustic-phonetic properties and the models for predicting voice quality categories from the speaker embeddings obtained from the \gls{fvae}. We then measure the prediction quality and estimate the Shapley values  to shed light on which information is stored where in the speaker embeddings.

\subsection{Factorized Variational Autoencoder}

\begin{figure}[t]
	\centering
	\begin{tikzpicture}[auto, node distance=0.5cm,>=latex', 		
	block/.style={
		rectangle,
		draw,
		text centered,	
	},
	]
	\node [] (input) {$\mathbf{X}$};
	\node[block, rounded corners, right = 1cm of input] (SE) {\stackanchor{Speaker}{Encoder}};
	\node[block, below = of input] (VTLP) {VTLP};
	\node[block, below = of VTLP] (IN) {\stackanchor{Instance}{Norm}};
	\node[block, rounded corners] at (IN.west -| SE.south) (CE) {\stackanchor{Content}{Encoder}};
	\node[block, right = of SE] (GAP) {GAP};
	\node[block, rounded corners, minimum height=1cm] at ($(VTLP.east -| GAP.east) + (1.5cm,0)$) (Decoder) {Decoder};
	\draw [->, thick] (input) -- (SE);
	\draw [->, thick] (input) -- (VTLP);
	\draw [->, thick] (VTLP) -- (IN);
	\draw [->, thick] (IN) -- (CE);
	\draw [->, thick] (SE) -- node[above] {$\mathbf{S}$} (GAP);
	\draw [->, thick] (GAP) |- ([yshift=0.25cm] Decoder.west);
	\node[above left = -0.2cm and 0.2cm of Decoder] (s) {$\bar{\mathbf{s}}$};
	\draw [->, thick] (CE) -- (CE.east -| GAP.south) |- ([yshift=-0.25cm] Decoder.west);
	\node at (Decoder.south -| s.south){$\mathbf{C}$};

	\node [right = of Decoder] (Output) {$\mathbf{\hat{X}}$};
	\draw[->,thick] (Decoder) -- (Output);
	\end{tikzpicture}
	\caption{Architecture of the FVAE}
	\label{Fig:FVAE}
	\vspace*{-0.4cm}
\end{figure}
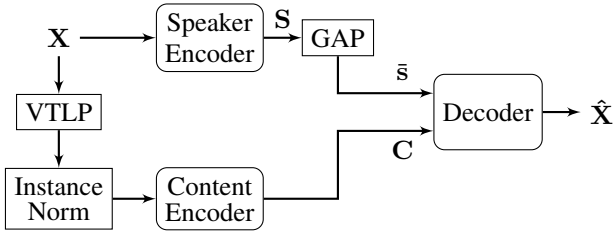

For our experiments we used  the \gls{fvae}, see Fig.~\ref{Fig:FVAE}, that disentangles content related from speaker related variations in the speech signal using two encoders, a content encoder and a speaker encoder. The input of the \gls{fvae} is the log-mel feature sequence $\vect{X}_\mathrm{mel}=(\vect{x}_1, \ldots , \vect{x}_T)$ of length $T$ frames, and the decoder output is the reconstructed sequence $\hat{\vect{X}}_\mathrm{mel}$ of the same length. Disentanglement is based on the hypothesis that content related variations of the speech signal change at a much faster rate than speaker related variations. Thus, the speaker embedding vector $\bar{\vect{s}}$ is obtained by \gls{gap} of the speaker encoder outputs $\vect{s}_t$, $t=1,\ldots , T$ over the time dimension. On the contrary, the content encoder emits content representations $\mathbf{C}= (\vect{c}_1, \ldots , \vect{c}_T)$ at frame rate. Disentanglement is actively encouraged by adding a \gls{cpc} loss on the speaker representation and an adversarial CPC loss on the content representation to the reconstruction training objective. The former encourages similarity of $\vect{s}_t$ and $\vect{s}_{t+\tau}$, where $\tau$ denotes the time lag (typically in the order of a second). To further obfuscate speaker information for the content encoder, \gls{vtlp} and instance normalization \cite{chou2019one} are applied to the input of the content encoder. 

The \gls{fvae} model is trained on the English data set Librispeech \cite{panayotov2015librispeech} and the decoder is further fine-tuned on the German data set LibriVoxDeEn \cite{LibriVoxDeEn} because it is used on a German data set as is described in the following. Disentanglement is achieved by the \gls{fvae} in a completely unsupervised fashion. Additionally, we employed a d-vector extractor for comparison purposes. It is a ResNet34 model that computes 256-dimensional speaker embeddings $\mathbf{x}$ from 80-dimensional log-mel spectrograms \cite{Cordlandwehr2023_diarization}. It is trained in a supervised manner using a speaker classification loss on the VoxCeleb speech corpus~\cite{nagrani2020voxceleb}.

\subsection{Prediction Models for Acoustic Properties}

The investigations on feature importance reported here, have been carried out on the \gls{nsc} corpus \cite{NSC}. This is a German data set consisting of $\approx 8\,\mathrm{h}$ scripted and $\approx 15\,\mathrm{h}$ of semi-spontaneous dialogues, with a total of $300$ speakers. Of those, 70\% of the speakers were used for training the prediction models, and the remaining 30\% for testing the models. We trained several prediction models on the speaker embeddings $\bar{\vect{s}}$. In a first set of experiments, we are interested in determining if speaker-specific acoustic-phonetic information is captured by  $\bar{\vect{s}}$. For this purpose, we utilized the extended \gls{eGeMAPS}, a set of 25 acoustic \glspl{lld} from which 88 functionals, including utterance-wise statistics such as mean and variance, are computed using openSMILE \cite{Eyben2016_diss, eyben2010opensmile}. From these, we have chosen the mean of the fundamental frequency $y_\mathrm{p}$, known to be speaker dependent, and the Hammarberg index $y_\mathrm{h}$, which is the ratio of the maximum power below $2000\,\mathrm{Hz}$ to the maximum power in the range $(2000 \,\mathrm{Hz} - 5000\,\mathrm{Hz}$) and it was used for emotion recognition \cite{tamarit2008spectral}. 
Further, we included the \gls{cpp} $y_{\mathrm{\gls{cpp}}}$ \cite{hillenbrand1994acoustic, drugman2014data} as additional speaker-dependent  feature. We also selected an acoustic feature that is considered to be more content than speaker dependent: the mean length of voiced segments $y_\mathrm{v}$.   

For each of these continuous-valued acoustic-phonetic signal properties, we trained various regression models to predict them from the extracted embeddings: \Gls{lr}, \gls{svr}, a \gls{mlp} with two hidden layers and two \glspl{vr}. \Gls{vr}-1 consists of an \gls{svr} and an \gls{lr} and \gls{vr}-2 of a \gls{svr} and a \gls{mlp}.

\subsection{Prediction Models for Voice Quality}

For the experiments on voice quality prediction, 300 speaker samples (174 female and 126 male speakers) of the semi-spontaneous dialogues of the \gls{nsc} data set, each about 30 seconds long, had been labeled by an expert phonetician with regards to presence or absence of certain voice quality categories. In this study, we concentrate on two of them, hoarseness (81 hoarse speakers), and nasality (59 nasal speakers). 
Although, only the semi-spontaneous dialogues were manually labeled, we used the same labels for the scripted dialogues, because they were spoken in the same session by the same speakers. This was considered justified because hoarseness and nasality are temporally persistent and thus carry over to the scripted dialogues if spoken by the same speaker. For each of the two voice quality categories we trained the following classifiers: \Gls{lda}, \gls{svc}  and a \gls{rf}.

\section{Results}
\label{sec:results}
\subsection{Acoustic Feature Prediction}

Table \ref{tab:linear_regression_models} shows the results for the prediction of the acoustic-phonetic signal properties from the speaker embedding vectors $\bar{\vect{s}}$ of the \gls{fvae} and $\vect{d}$ of the d-vector extractor. Results are given in terms of coefficient of determination between the values $y$ computed by openSMILE from the input speech $\vect{X}$ and their estimates $\hat{y}$ obtained from the predictor:
\begin{equation}
    R^2(y, \hat{y}) = 1 - \frac{\ExpOp[(y - \hat{y})^2]}{\ExpOp[\left(y - \ExpOp[y]\right)^2]} \, , 
\end{equation}
whose optimal value is equal to $1$, which results  if $\hat{y}$ completely matches $y$. If the result is equal to $0$, then the prediction is no better than the mean value of $y$. It can be observed, that the performance of the different regression models is quite comparable. Looking at the predictions from the d-vectors, it is obvious, that the  features $y_\mathrm{p}$ and  $y_{\mathrm{\gls{cpp}}}$, which are considered speaker-dependent, can be much better predicted than  $y_\mathrm{v}$, which was considered speaker-independent. 
Interestingly, the prediction results  from $\bar{\vect{s}}$  for $y_\mathrm{p}$, $y_{\mathrm{\gls{cpp}}}$ and $y_\mathrm{v}$ are quite similar to those from $\vect{d}$. This is an indication that the assumption underlying the computation of $\bar{\vect{s}}$ in the \gls{fvae}, that temporally stable signal properties carry speaker information, is justified, as the computation of $\vect{d}$, which explicitly uses  speaker information, leads to similar results. This is not true for $y_\mathrm{h}$, which leads to some doubts if $y_\mathrm{h}$ is indeed speaker-dependent.

\begin{figure}[!t]
	\input{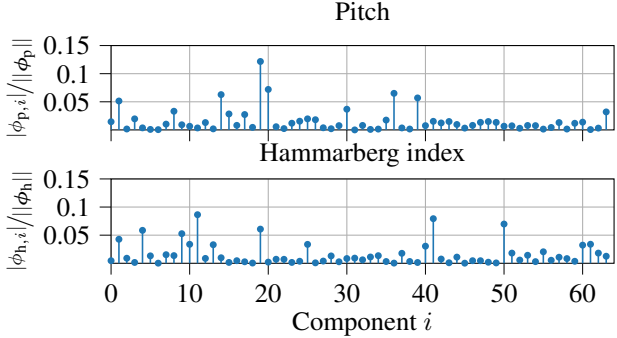}
	\vspace{-0.75cm}
	\caption{Normalized absolute values of the estimated Shapley values to explain the model output of the \gls{vr}-2 for pitch and \gls{vr}-2 for Hammarberg index.}
	\label{Fig:projection_vectors}
	\vspace{-0.4cm}
\end{figure}

To identify which components of the speaker vector carry the information 
about a certain acoustic-phonetic property, we used the trained \gls{vr}-2 models and estimated the Shapley values of Eq.~\eqref{Eq:Additive_feature_attribution_methods} with kernel\gls{shap} \cite{lundberg2017unified}. The Shapley values of 3000 random utterances of the training set are estimated. Figure \ref{Fig:projection_vectors} displays the mean absolute Shapley values for each component of the 64-dimensional speaker vector $\bar{\vect{s}}$ for the prediction of the mean pitch and the Hammarberg index, normalized such that its overall sum is 1. What is most striking from this result, is, that the information about the investigated acoustic-phonetic property is pretty much spread out over the components of the speaker embedding vector. The speaker vector lacks in compactness and modularity \cite{carbonneau2022_disentanglement, Eastwood}, where compactness refers to the dispersion of information across dimensions and modularity assesing the encapsulation of acoustic signal properties within a single dimension.

\begin{table}[t]
\centering
\begin{tabular}{
c
p{0.35cm}>{\centering}
p{0.45cm}>{\centering}
p{0.35cm}>{\centering}
p{0.45cm}>{\centering}
p{0.45cm}>{\centering}
p{0.45cm}>{\centering}
p{0.35cm}
c}
\toprule 
\multirow{3}{*}{Model} & \multicolumn{8}{c}{$R^2(y, \hat{y}) $ }\\
\cmidrule{2-9}
 & \multicolumn{2}{c}{\multirow{1}{*}{$y_\mathrm{p}$}} & \multicolumn{2}{c}{\multirow{1}{*}{$y_\mathrm{h}$}}  & \multicolumn{2}{c}{\multirow{1}{*}{$y_\mathrm{\gls{cpp}}$}} & \multicolumn{2}{c}{\multirow{1}{*}{$y_\mathrm{v}$}}\\
\cmidrule(r){2-3} \cmidrule(lr){4-5} \cmidrule(lr){6-7} \cmidrule(lr){8-9}
& $\bar{\mathbf{s}}$ & $\mathbf{d}$ & $\bar{\mathbf{s}}$ & $\mathbf{d}$ & $\bar{\mathbf{s}}$ & $\mathbf{d}$  & $\bar{\mathbf{s}}$ & $\mathbf{d}$  \\
\midrule
\acrshort{lr}  &  .95 & .93 & .83 & .54 & .74 & .69 & .36 & .46\\
\acrshort{svr} &  .96 & .93 & .79 & .51 & .66 & .67 & .32 & .46\\
\acrshort{mlp} &  .96 & .$\mathbf{95}$ & .83 & .48 &  .73 & .64 & .36 & .43\\
\acrshort{vr}-1 & $.\mathbf{97}$ & $.\mathbf{95}$ & $.\mathbf{85}$ & $.\mathbf{57}$ & $.\mathbf{75}$ & $.\mathbf{72}$ & $.\mathbf{42}$ & $.\mathbf{51}$\\
\acrshort{vr}-2 & $.\mathbf{97}$ & $.\mathbf{95}$ & .84 & .54 & .74 & .70 & $.\mathbf{42}$ & $.\mathbf{51}$\\
\bottomrule
\end{tabular}
\caption{Performance of various regression models trained on speaker embedding $\bar{\mathbf{s}}$ or d-vector $\mathbf{d}$ to predict certain acoustic-phonetic signal properties on the \gls{nsc} corpus.}
\label{tab:linear_regression_models}
\vspace{-0.38cm}
\end{table}

\subsection{Voice Quality Prediction}

In light of the very limited training set size, we used 4-fold cross validation for the prediction of the binary voice quality features. We divided the folds in such a way, that there were the same number of  speakers with the presence and with the absence of the voice quality category in each set. Table \ref{tab:accuracy_clf} shows the F1-score  of the classifiers, averaged over all folds. 
For comparison purposes we also added the performance of a gender classifier trained on the same input features.
\begin{table}[t]
\centering
\begin{tabular}{ccccccc}
\toprule
\multirow{3}{*}{Model}& \multicolumn{6}{c}{F1-Score$\,$[\%]}\\
\cmidrule{2-7}
  & \multicolumn{2}{c}{Gender} & \multicolumn{2}{c}{Nasality} & \multicolumn{2}{c}{Hoarseness}\\
  \cmidrule(r){2-3} \cmidrule(lr){4-5} \cmidrule(lr){6-7}
  & $\bar{\mathbf{s}}$ & $\mathbf{d}$ & $\bar{\mathbf{s}}$ & $\mathbf{d}$& $\bar{\mathbf{s}}$ & $\mathbf{d}$\\
\midrule
\acrshort{svc} & 99 & $\mathbf{100}$ & $\mathbf{63}$ & $\mathbf{62}$ & $\mathbf{56}$ & $\mathbf{59}$\\
\acrshort{lda} & 99 & $\mathbf{100}$ & $\mathbf{63}$ & 59 & 55 & 56\\
\acrshort{rf} & 99 & 99 & 58 & 61 & $\mathbf{56}$ & 56\\
\bottomrule
\end{tabular}
\caption{F1-score of classifiers to predict voice qualities by using the information of the speaker embedding $\bar{\mathbf{s}}$ or d-vector $\mathbf{d}$. \gls{nsc} corpus is used for training.}
\label{tab:accuracy_clf}
\vspace{-0.4cm}
\end{table}
The results show that all classifiers, be it with $\bar{\vect{s}}$ or $\vect{d}$ at its input, were able to very reliably predict the gender of the speaker, whereas the voice quality classifiers are hardly  better than guessing. To put these results into perspective, in \cite{Wiechmann_2023} experiments are reported on predicting hoarseness using six carefully selected acoustic-phonetic signal properties (F0, intensity, \gls{cpp}, F2 and spectral slope measures) that were computed on certain vowel realizations, which were found by forced aligned segments of the input speech. Since the same data set was used as in our study, their results, which had been obtained  with a \gls{rf} classifier, can be readily compared with the results obtained here. They reported an F1-score  of 0.87, which is much higher. This clearly shows that a global speaker embedding, as it is computed by the \gls{fvae} or the d-vector extractor,  does not capture voice quality aspects well, despite the fact that we here tested voice quality categories that are considered to be temporally persistent and speaker-dependent.


\section {Conclusions and Outlook}
\label{sec:conclusion}

In this work, we examined the ability of the speaker embedding vectors of the \gls{fvae} and a d-vector extractor to capture acoustic-phonetic signal properties and voice quality aspects. It was observed that the tested acoustic features that are known to be speaker-specific, could be fairly well predicted from the speaker embedding vector, while the more abstract voice quality categories hoarseness and nasality could not. This result asks for a more elaborated way of  estimating the speaker embedding. 
Future work should replace the \gls{gap} in the \gls{fvae} by more sophisticated pooling operations, such as attention, to focus on those segments of the speech that are most informative of a certain voice quality feature. 
Further, we employed the concept of Shapley values to determine the attribution of embedding vector components to the prediction results. It was shown that the acoustic signal properties are not disentangled in the embedding vector space, which would make it difficult to manipulate one 
without affecting the other. Nevertheless, the findings reported here give important insights into the properties of speaker embedding vectors.

\section*{Acknowledgement}
\noindent Funded by the Deutsche Forschungsgemeinschaft (DFG, German Research Foundation): TRR 318/1 2021- 438445824.

\balance
\bibliographystyle{ieeetr}
\bibliography{example}

\end{document}